\documentclass[11pt,a4paper]{scrartcl}

\usepackage[utf8]{inputenc}
\usepackage{lmodern}
\usepackage[british]{babel}
\usepackage{amsmath}
\usepackage{amssymb}
\usepackage{amsfonts}
\usepackage{color}
\usepackage{url}
\usepackage{hyperref}
\usepackage{bbm}
\usepackage{slashed}
\usepackage{graphicx}
\usepackage{caption}
\usepackage{subcaption}
\usepackage{simplewick}
\usepackage{verbatim}
\usepackage[dvipsnames]{xcolor}
\usepackage{empheq}
\usepackage{textcomp}
\usepackage{authblk}
\DeclareOldFontCommand{\bf}{\normalfont\bfseries}{\mathbf}

\newcommand{\arxiv}[1]{arXiv:\,\href{http://arxiv.org/abs/#1}{{\texttt #1}}}

\begin{document}
	\title{Measurement of the mass anomalous dimension of near-conformal adjoint QCD with the gradient flow}
	\author[1,2]{Camilo Lopez\thanks{camilo.lopez@uni-jena.de}}
	\author[1]{Georg Bergner\thanks{georg.bergner@uni-jena.de}}
	\author[3]{Istvan Montvay\thanks{montvay@mail.desy.de}}
	\author[4]{Stefano Piemonte\thanks{stefano.piemonte@ur.de}}
	\affil[1]{University of Jena, Institute for Theoretical Physics, 
		Max-Wien-Platz 1, D-07743 Jena, Germany}
	\affil[2]{Department of Physics, University of Colorado, Boulder, Colorado 80309, United States}
	\affil[3]{Deutsches Elektronen-Synchrotron DESY, Notkestr.~85, D-22607
		Hamburg, Germany}
	\affil[4]{University of Regensburg, Institute for Theoretical Physics, 
		Universit\"atsstr.~31, D-93040 Regensburg, Germany}
	\maketitle
	\begin{abstract}

		The mass anomalous dimension is determined in SU(2) gauge theory coupled to $N_f$ fermions in the adjoint representation for $N_f =  2$, $3/2$, $1$ and $1/2$, where
		half-integer flavor numbers correspond to Majorana fermions. The numerical method is based on gradient flow.

		The results show near-conformal behavior for $N_f = 2$, $3/2$ and $1$.
		Particular emphasis is placed on $N_f = 2$ which is relevant for a strongly interacting
		extension of the Standard Model and has been studied in several previous investigations.
		We check whether the method is able to resolve discrepancies in earlier results for this theory.
		Overall, the method based on gradient flow delivers reliable results in qualitative agreement with previously known numerical data.
	\end{abstract}

  The landscape of strongly coupled gauge theories with  different 
 matter contents has been the subject of many recent investigations. They are 
 motivated by the search for a strongly interacting completion of the 
 Standard Model and by an improvement our general understanding of strong interactions, 
 see reviews \cite{Svetitsky:2017xqk, Appelquist:2016viq}. 
 A QCD-like behavior of these theories implies 
 a running of the coupling from asymptotic freedom corresponding to a Gaussian 
 ultra violet (UV) fixed point towards confinement in the infrared 
 (IR). Confinement is due to an exponential growth of the strong coupling 
 implying non-perturbative phenomena like the existence of a mass gap 
 even in the limit where all mass parameters of the matter fields are zero. 
 The remormalization group (RG) flow of QCD from the perturbative running 
 towards the strongly coupled infrared has been a subject of several studies.
 
 Increasing the matter content of the theory leads to a screening effect, 
 which will change this scenario. Eventually, asymptotic freedom is lost
 and the RG flow is changed towards a more QED-like form. In between 
 the QCD-like running and the theories with no asymptotic freedom
 we expect the existence of the so-called conformal window. The matter fields
 in this regime screen the running in the IR, but asymptotic freedom
 is still preserved in the UV. Consequently, the strong coupling increases until 
 it reaches a fixed point in the IR. The RG flow close to the fixed point 
 is characterized by universal quantities like mass ratios and anomalous 
 dimensions of certain operators. 
 
 Once the lower boundary of the conformal window is reached, the slow running of 
 the gauge coupling provides interesting scenarios for model building. 
 A large scale separation appears while the theory is still at strong 
 couplings. This could explain an enhancement of certain operators 
 in the low energy effective theory if the anomalous dimensions are 
 large enough. In addition, a light dilaton-like scalar field can exist. 
 From the perspective of model building, a smaller 
 number of matter fields is preferred. This has lead to consider theories 
 with fermions in higher representation which require a smaller 
 number of fields to reach a near-conformal behavior. One of the most prominent 
 examples of such theories is the SU(2) 
 gauge theory coupled to two adjoint Dirac fermions
 \cite{Bergner:2017gzw,Rantaharju:2015yva,
 	DeGrand:2011qd,DelDebbio:2010hx,
 	Catterall:2008qk}. 
 
 The motivation for our investigation is the exploration of the general 
 theory space for possible realizations of strongly interacting gauge theories. 
 We investigate SU(2) gauge theories with different number of fermions in the 
 adjoint representation (adjoint QCD) in order to characterize 
 the lower boundary of the conformal window. We denote by $N_f$ the number 
 of Dirac fermions. Half integer values of $N_f$ correspond to $2N_f$ Majorana 
 fermions. The important question is not only whether a fixed point exists 
 for these theories. Going into more detail,
 we focus here on the mass anomalous dimension as one of 
 the most important characteristics of the infrared fixed point (IRFP). Our 
 main results are related to the $N_f=2$ case, since the conformal behavior of this theory 
 is indicated by several previous numerical lattice studies. Therefore, 
 it is an ideal starting point to test new methods determining
 the RG flow close to the fixed point. In addition, our aim is to resolve the observed
 discrepancies in the previously obtained values for the mass anomalous dimension.
 
 In earlier studies we have also considered $N_f=1$, $N_f=3/2$, as well as 
 $N_f=1/2$ \cite{Bergner:2015adz,Athenodorou:2014eua,Bergner:2016hip}. 
 $N_f=1/2$ serves as a cross-check since it corresponds to $\mathcal{N}=1$ 
 supersymmetric Yang-Mills theory and should be outside of the conformal 
 window. These studies were intended to investigate how the mass anomalous 
 dimension increases towards the lower end of the conformal window until 
 the fixed point disappears. 
 In theories with good indications for a conformal fixed point only quite 
 small mass anomalous dimensions have been observed.
 Therefore, more detailed studies could show if much larger values could 
 be realized at all.
 
 The investigations of the IRFP on the lattice is hampered by several 
 difficulties. The RG flow is restricted in the ultraviolet by the finite 
 lattice spacing and in the infrared by the finite mass parameters and by 
 the finite volume. This restricts the range of scales for the investigations 
 and leads to deformations of the flow which are hard to quantify.
 
 The methods we have applied in our earlier studies for the determination 
 of the mass anomalous dimension $\gamma_{*}$ are based on the scaling 
 of the mass spectrum \cite{Bergner:2016hip} and the mode 
 number \cite{Patella:2012da}. The bound state masses 
 $M$ are predicted to scale with the fermion mass $m$ according to 
 $M\approx m^{1/(1+\gamma_{*})}$. The mode number computed from the eigenvalues 
 of the Dirac operator is predicted to scale in a certain regime between 
 the infrared, dominated by the mass and the volume corrections, 
 and the ultraviolet, dominated by lattice artifacts and 
 perturbative running. We have found that the anomalous dimension varies 
 significantly with the bare gauge coupling $\beta$. Recently a method 
 has been proposed to determine $\gamma_{*}$, which captures the detailed features 
 of the RG running and seems to be able to resolve the discrepancy among 
 different $\beta$ values 
 \cite{Carosso:2018bmz}. 
 The different bare parameters can eventually be extrapolated to a convergent flow 
 towards a single fixed point value. In this work we test whether this method 
 provides values consistent with our previous determinations and whether it 
 might help to resolve the discrepancies.

\section{The anomalous dimension from the gradient flow}

In this section we briefly review the method for the determination of the mass anomalous dimension first derived in \cite{Carosso:2018bmz}. The fixed point value  $\gamma_{*}$ of this quantity is extrapolated from the RG flow. A RG step consists of a coarse-graining and a scale transformation (dilation). In Refs.~\cite{Carosso:2019qpb,Carosso:2018bmz}, it has been proposed that the gradient flow corresponds (asymptotically at large flow times) to the coarse-graining step. This relation arises from the fact that, for a given flow time $t$, the gradient flow kernel smears the fields over a (Gaussian) radius $\sqrt{8t}$. In momentum space this is equivalent to imposing a smooth UV-cutoff. In the following we discuss how the effect of the dilation can be included, so that we can explore the RG flow and its fixed points from the gradient flow.

A RG step with parameter $b>1$ scales the lattice spacing according to $a\to a^{\prime}=b\,a$, and changes the couplings as $g\to g^{\prime}$, $m\to m^{\prime}$. Generic fields $\phi$ with canonical dimension $d_\phi$ are scaled according to their scaling dimension, which includes the anomalous $\eta$.
A generic two-point function transforms as
\begin{align}\label{gf1}
\langle\mathcal{O}(0)\mathcal{O}(x_{0})\rangle_{g,m}=b^{-2(d_{\mathcal{O}}+\gamma_{\mathcal{O}})}\langle\mathcal{O}(0)\mathcal{O}(x_{0}/b)\rangle_{g',m'},
\end{align}

where the operator $\mathcal{O}(x_0)\equiv\mathcal{O}[\phi](x_{0})$ is a lattice interpolator, which is in general a monomial of gauge or fermion fields of order $n_{\mathcal{O}}$\footnote{Eq.~\ref{gf1} holds as long as $\mathcal{O}$ are scaling operators, i.e.\ if they are eigenstates of the linearized RG equations.}. The scaling dimension $\Delta_{\mathcal{O}}$ includes the classical ($d_{\mathcal{O}}$) and anomalous dimension ($\gamma_{\mathcal{O}}$). 

Although the gradient flow doesn't re-scale the smeared fields, we can include a renormalization factor by relating $t$ and $b$. At large $t$ it is expected that\footnote{We stress that this is only valid at large $t$, since as $t\to0$ it should hold that $b\to 1$.} $b\propto\sqrt{t}$
\begin{align*}
	\phi_{b}(x_{0}/b)=b^{\Delta_{\phi}}\phi_{t}(x_{0})=b^{d_{\phi}+\eta/2}\phi_{t}(x_{0}).
\end{align*}

Here the subscript $b$ labels the blocked fields after the RG step\footnote{The Monte Carlo Renormalization Group method states that the right hand side of Eq.\ref{gf1} can be written as a correlator of blocked fields with respect to the UV action.}. Assuming that $x_{0}$ is large enough to both neglect the re-scaling of the distances on the lattice and avoid the overlapping of the operators as $t$ increases, we get

\begin{align}\label{ratio1}
\frac{\langle\mathcal{O}_{t}(0)\mathcal{O}_{t}(x_{0})\rangle}{\langle\mathcal{O}(0)\mathcal{O}(x_{0})\rangle}= t
^{\Delta_{\mathcal{O}}-n_{\mathcal{O}}\Delta_{\phi}}\; ,
\end{align}
The index $t$ indicates interpolators consisting of fields at flow time $t$. Note that it is assumed that the $x_0$ dependence cancels between numerator and denominator for large enough $x_0$. Moreover, the ratio doesn't require us to know the exact relation $b\propto \sqrt{t}$, if we stay away from small $t$.

In order to remove the dependence on the scaling dimension of the field $\phi$, a ratio with operators ($\mathcal{V}$) of scaling dimension $\Delta_{\mathcal{V}}=0$ is added. This is fulfilled if $\mathcal{V}$ is a conserved current. For simplicity we assume that it has the same field content as $\mathcal{O}$ ($n_{\mathcal{O}}=n_{\mathcal{V}}$ and $d_{\mathcal{O}}=d_{\mathcal{V}}$). This leads to 
\begin{align}\label{RT1}
\frac{\langle\mathcal{O}_{t}(0)\mathcal{O}_{t}(x_{0})\rangle}{\langle\mathcal{O}(0)\mathcal{O}(x_{0})\rangle}\frac{\langle\mathcal{V}(0)\mathcal{V}(x_{0})\rangle}{\langle\mathcal{V}_{t}(0)\mathcal{V}_{t}(x_{0})\rangle}
= t^{\gamma_{\mathcal{O}}}\; .
\end{align}

The computation of correlators with both interpolators consisting of fields at $t$ requires the integration of the computationally expensive adjoint fermion flow equation. As a technical simplification one considers the correlator with only the sink term at flow time $t$,
\begin{align}\label{RT}
\mathcal{R}_{\mathcal{O}}(t,x_{0})&=\frac{\langle\mathcal{O}(0)\mathcal{O}_{t}(x_{0})\rangle}{\langle\mathcal{O}(0)\mathcal{O}(x_{0})\rangle}\frac{\langle\mathcal{V}(0)\mathcal{V}(x_{0})\rangle}{\langle\mathcal{V}(0)\mathcal{V}_{t}(x_{0})\rangle}
= t^{\gamma_{\mathcal{O}}/2}\; .
\end{align}
The corrections are of order $\mathcal{O}(a\sqrt{t}/x_{0})$ and, following  \cite{Carosso:2018bmz}, they are neglected in the analysis. The final formula for the  scaling of $\gamma_{\mathcal{O}}$ with the energy scale is ($\bar{t}=(t_1+t_2)/2$)
\begin{align}\label{GammaO}
\gamma_{\mathcal{O}}(\bar{t})=\frac{\log(\mathcal{R}_{\mathcal{O}}(t_{1})/\mathcal{R}_{\mathcal{O}}(t_{2}))}{\log{(\sqrt{t_{1}}/\sqrt{t_{2}})}}.
\end{align}
It is assumed that $\mathcal{R}$ doesn't depend for large enough $x_0$ and hence $\mathcal{R}(t,x_0)$ should approach a constant. 

In the measurements on the lattice several deformations of the running have to be taken into account. In a vicinity of the IRFP the relevant deformations are the finite mass and the finite volume. This requires especial care, since Eq. \eqref{RT1} only holds on the critical surface, i.e. at infinite volume and zero mass. In addition, the effects of the lattice discretization will have a significant impact on the observed renormalization group flow towards the fixed point on the lattice. In practice, the effective anomalous dimension at the energy scale $\mu$ is represented using $\mu=1/\sqrt{8\bar t}$. 

\section{The mass anomalous dimension of adjoint QCD } 
The continuum action of the SU(2) gauge theories with fermions in the adjoint representation reads
\begin{equation}
S = \int d^4 x \left\{\frac{1}{4} (F_{\mu\nu}^a F_{\mu\nu}^a) + \frac{1}{2} \sum_{i=1}^{2 N_f}\bar{\lambda}_a^i \gamma^\mu D^{ab}_\mu \lambda_b^i\right\}\,.
\end{equation}
In this formulation, the $2N_f$ fermion fields $\lambda^i$ fulfill the Majorana condition, but for integer $N_f$ it can 
be straightforwardly converted into a theory with $N_f$ Dirac fermions.
The covariant derivative acts in the adjoint representation is defined as
\begin{equation}
D^{ab}_\mu \lambda_b = \partial_\mu \lambda_a + i g A_\mu^c (T_c^A)^{ab} \lambda_b\,,
\end{equation}
where $T_c^A$ are the Lie algebra generators as given by the structure constants. On the lattice, we employ a tree-level Symanzik improved gauge action and stout-smeared Wilson fermions.
The simulation parameters are summarized in Table \ref{tablenf2}.
Note that a sign problem can arise for odd number of fermions, but our simulations are performed in a range of fermion masses where it can be neglected.

We measured the anomalous dimension of the pseudo-scalar operator, which is related to the mass anomalous dimension as $\gamma_{m}=-\gamma_{PS}$, in adjoint QCD with $N_{f}=1/2, 1, 3/2$, and $2$. As mentioned earlier, our emphasis lies on the two-flavour case, since most other studies consider this theory. 
SYM is the only of these theories, which is known to lie outside the conformal window. The interpolator of the pseudo-scalar operator is $\bar\lambda\gamma_{5}\lambda$. 

The natural choice of $\mathcal{V}$ in case of Wilson fermions is the vector current, as it is the simplest conserved current on the lattice. It is however important to notice, that the local (continuum) vector current
\begin{align}\label{localv}
\mathcal{V}_{\mu}^a=\bar\lambda\frac{\tau^{a}}{2} \gamma_{\mu}\lambda
\end{align}
is not exactly conserved at non-zero lattice spacing in contrast to the point-split lattice vector current
\begin{align}
\label{nonlocalv}
\tilde{\mathcal{V}}_{\mu}^{a}=\frac{1}{2}\left[ \bar\lambda(x)(\gamma_{\mu}-1)U_{\mu}(x)\frac{\tau^{a}}{2}\lambda(x+a\hat{\mu})+ \bar\lambda(x+a\hat{\mu})(\gamma_{\mu}+1)U_{\mu}^{\dagger}(x)\frac{\tau^{a}}{2}\lambda(x) \right].
\end{align}
Here $\tau^{a}$ are the Pauli matrices acting on the flavor indices of the two Dirac spinor fields $\lambda$. If the local current $\mathcal{V}^a_{\mu}$ is used in the calculations, an additional dependence on the renormalization factor $Z_{\mathcal{V}}\neq 1$ remains. In particular, lattice artifacts are expected to introduce a spurious running of $Z_{\mathcal{V}}$ on the scale $\mu$, and we must ensure numerically that we are exploring a scaling region where such effects are under control. We actually perform the measurements of $\gamma_{PS}$ with both currents for the $N_f = 2$ case in order to ensure that the running in Eq.\eqref{RT} is dominated by the mass anomalous dimension and to cross check the results

The result of the computation is an effective anomalous dimension $\gamma_{m}$ as a function of the flow scale $\bar{t}$. For an IR conformal theory, it is expected that the running of $\gamma_{m}$ towards the IR stops at a certain energy scale and $\gamma_{m}$ converges to the fixed point value $\gamma_{*}$.
In practice, however, it is impossible to consider arbitrary large flow times due to non-zero size and finite mass effects in the IR. 
The value of $\gamma_{*}$ has to be extrapolated from intermediate flow times, which leads to the usual windowing problem. 
The main results are obtained at very small masses, like in \cite{Carosso:2018bmz}, which means that the finite volume effects are the dominant deformations of the flow.

\subsection{Adjoint QCD with two Dirac flavours}
We start our studies with the case of $N_{f}=2$ since in this case there are several previous studies available for comparison. All studies indicate the existence of an IRFP, but the anomalous dimension varies significantly. Our aim is check whether the gradient flow method is able to resolve the discrepancies. We simulated four different $\beta$ values (see Table \ref{tablenf2}): $\beta=1.5$ is close the bulk transition and we have seen that at this point we are able to reproduce the results in \cite{DelDebbio:2010hx,Patella:2012da}. In our previous investigations \cite{Bergner:2016hip}, we have seen that $\beta=1.7$ leads to quite different results and we complement the investigations with $\beta=1.6$ and $\beta=2.25$. Two different volumes, $V=24^{3}\times 64$ and $V=32^{3}\times 64$, are considered to estimate finite volume effects. In all cases we have measured $\mathcal{O}(100)$ well separated configurations. The gradient flow-times are in the range $1\leq t/a^2\leq 9$ since we have seen significant finite volume effects starting at $t/a^2=10$. We use the following scaling relation based on data at the volumes $L$ and $sL$
\begin{align}
\label{volumescaling}
\mathcal{R}_{\mathcal{O}}(g,s^{4}t,s^{2}L)=\mathcal{R}_{\mathcal{O}}(g,s^{2}t,sL)+s^{-\gamma_{\mathcal{O}}}\left( \mathcal{R}_{\mathcal{O}}(g,t,sL)-\mathcal{R}_{\mathcal{O}}(g,t,L) \right)+O(g'-g),
\end{align}
in order to obtain data in a third, even larger, effective volume $V=s^{2}L\times 64$ \cite{Carosso:2018bmz}, for the present parameters $s=32/24$, i.e.\ $L_{S}\sim 42$. This is done in order to reduce finite volume effects.

\subsubsection{Local and point-split currents}
\begin{figure}[htbp!]
	\centering
	\begin{minipage}[t]{0.49\textwidth}
		\includegraphics[width=\textwidth]{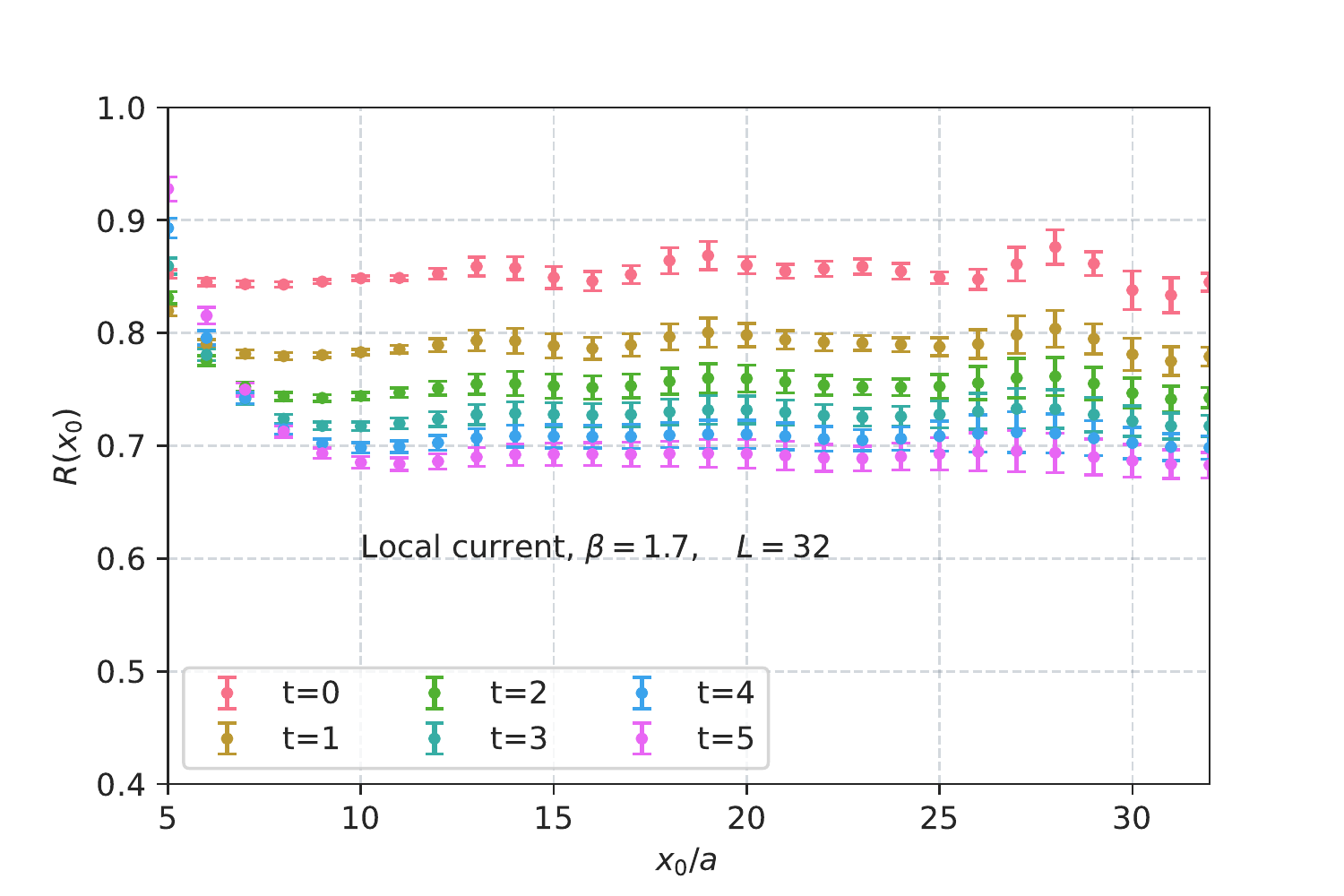}
		\caption{\footnotesize  $\mathcal{R}_{PS}(x_{0})$ from local vector current.}
	\end{minipage}
	\hfill
	\begin{minipage}[t]{0.49\textwidth}
		\includegraphics[width=\textwidth]{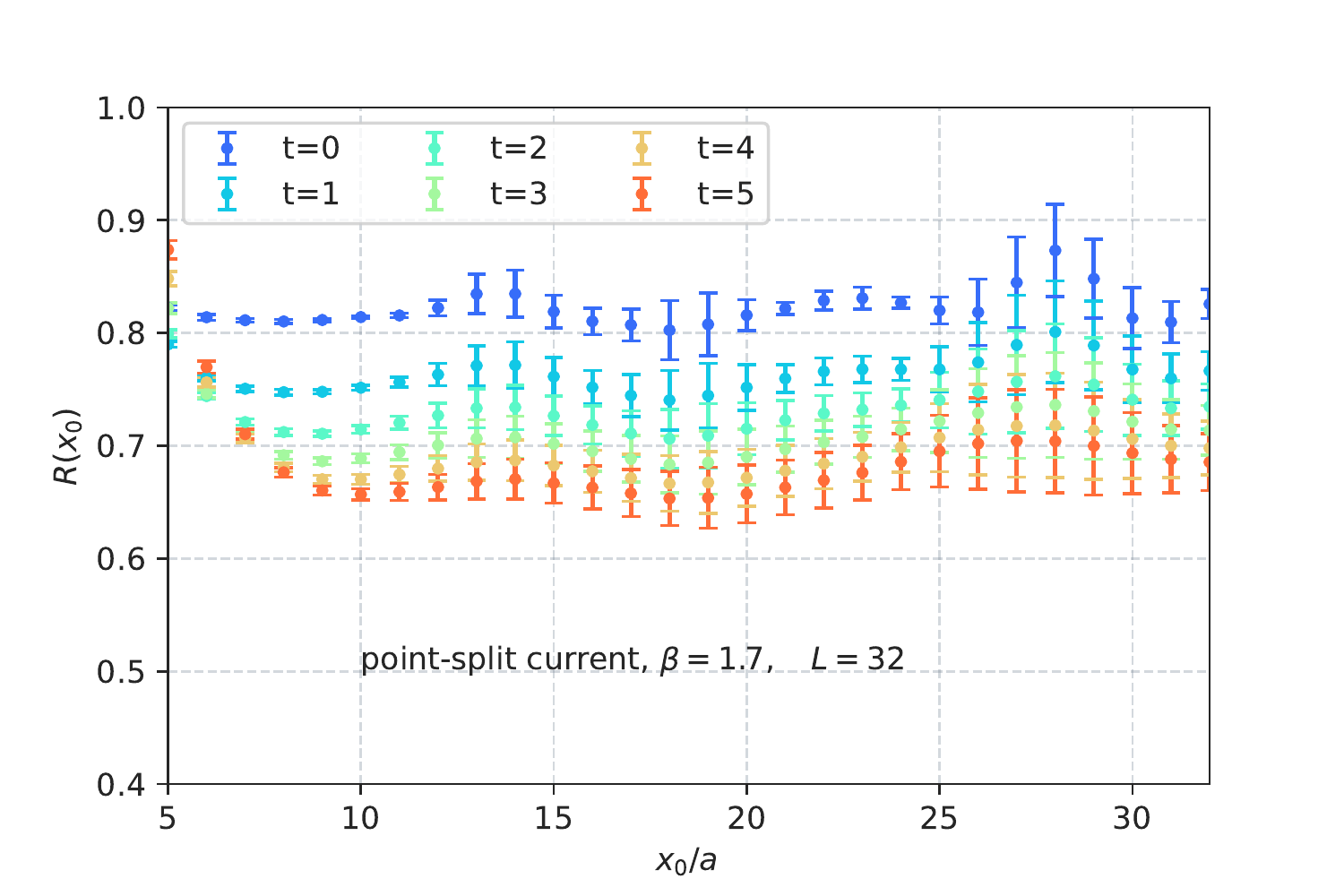}
		\caption{ \footnotesize $\mathcal{R}_{PS}(x_{0})$ from point-split vector current.}
		\label{Rplot1}
	\end{minipage}
	
	\begin{minipage}[t]{0.49\textwidth}
		\includegraphics[width=\textwidth]{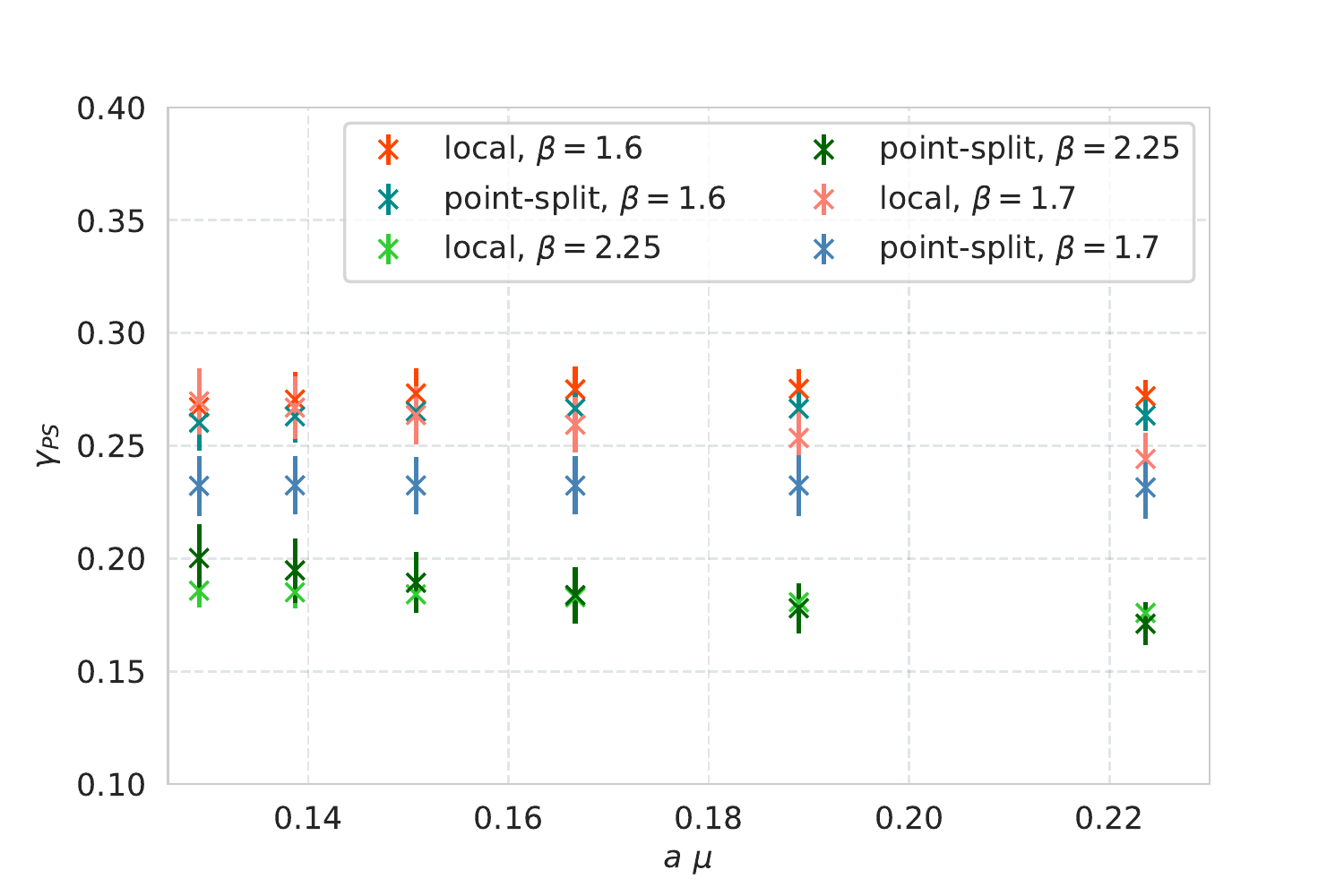}		
		\caption{\footnotesize Comparison of $\gamma_{PS}$ from local and point-split vector currents. $L=32$}
		\label{gammal32b17}
	\end{minipage}\hfill
	\begin{minipage}[t]{0.49\textwidth}
		\includegraphics[width=\textwidth]{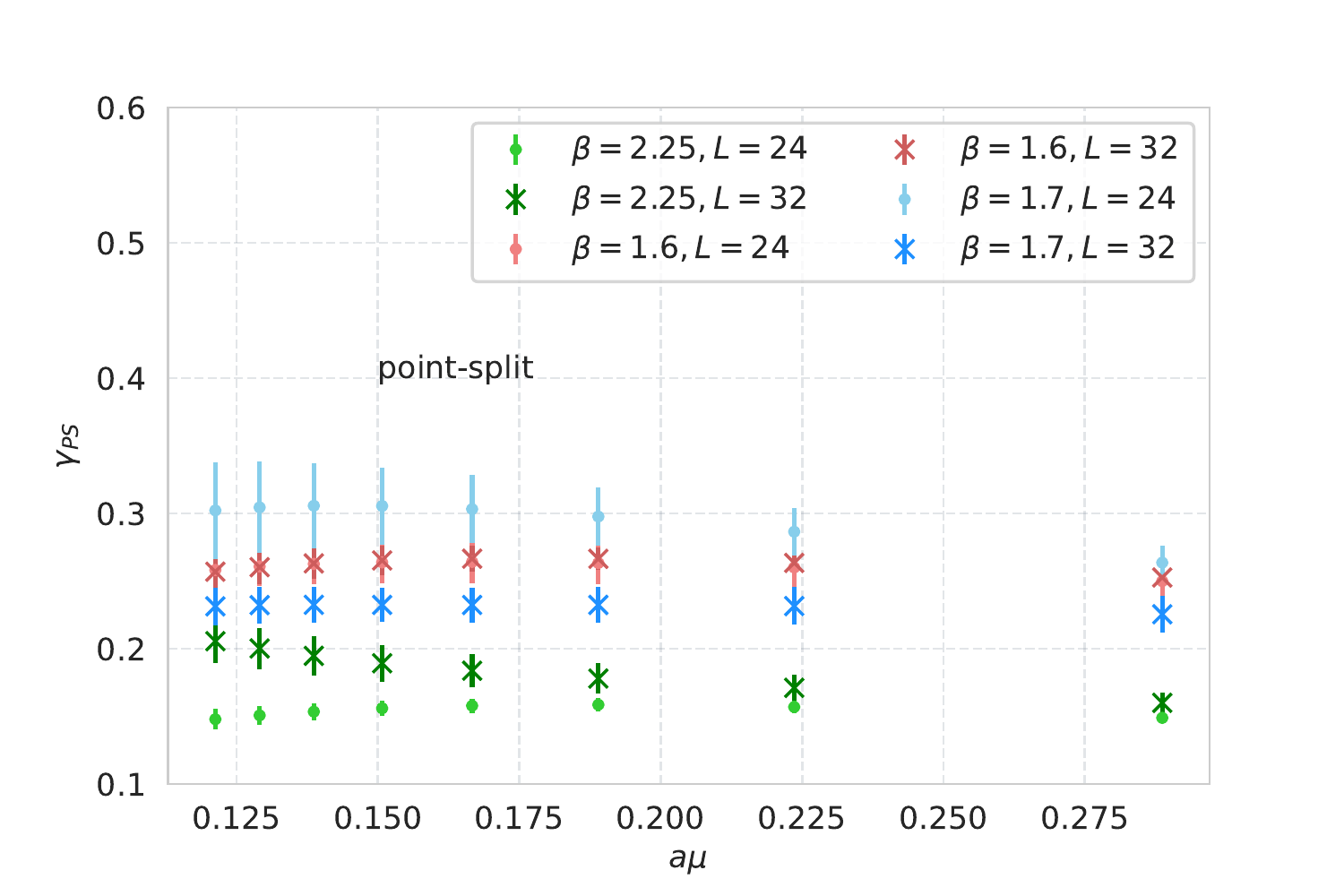}
		\caption{\footnotesize Volume dependence of $\gamma_{PS}$ computed with the point-split vector current.}
		\label{nf2finiteL}
	\end{minipage}
\end{figure}

The first part of the analysis is to determine the ratios $\mathcal{R}_{PS}(t,x_{0})$, which are easily obtained from the vector and pseudo-scalar operators. In Fig.~\ref{Rplot1} we show the ratio $\mathcal{R}_{PS}$ as a function of $x_{0}$ at different flow-times. 

We compare the results from both the local and point-split vector currents. We see that, for the local current, $\mathcal{R}_{PS}(x_{0})$ clearly converges to an almost constant plateau. For the point-split current, $\mathcal{R}_{PS}$ has larger fluctuations. This general observation is consistent for all the volumes and $\beta$ values we considered. As a measure of the asymptotic value of $\mathcal{R}_{PS}(t)$ at large distances in Eq. \eqref{GammaO}, we take the average of $\mathcal{R}$ over the interval $x_{0}\in[15,20]$. 
In Fig.~\ref{gammal32b17} we present as an example a direct comparison of $\gamma_{PS}(\mu)$ for our three larger $\beta$ at lattice size $L=32$, for the two kinds of vector currents. The results are compatible, but the uncertainties are larger in the point-split current case. Overall the renormalization factor of $\mathcal{V}_{\mu}$ seems,  as a first approximation, to be negligible for the final result.
The local current is more stable but comes with an additional systematic uncertainty since it is not strictly conserved. The point-split vector current, on the other hand, is conserved but leads to larger statistical errors. For $N_f=2$, we have employed both vector currents in the analysis at $\beta=1.6$, $1.7$ and $2.25$. For $N_f=3/2$ and $N_f=1$, we have focused on the results obtained from the point-split current.

\begin{table}[h]
	\centering
	\scalebox{0.9}{\begin{tabular}{ | c | c | c | c | c | c | }
			\hline
			$N_{f}$&$L_{S}$&$L_{T}$&$\beta$&$\kappa$& $am_{\mathrm{PCAC}}$\\ \hline
			2&24 &64 &1.5 &0.1350 & 0.03136(15) \\
			2&32 &64 &1.5 &0.1350 & 0.030414(45) \\\hline
			2&24 &64 &1.6 &0.1340 & 0.00869(23) \\
			2&32 &64 &1.6 &0.1340 & 0.00908(13) \\\hline
			2&24 &64 &1.7 &0.1328 & 0.00878(18) \\
			2&32 &64 &1.7 &0.1328 & 0.00894(13) \\\hline
			2&24 &48 &2.25 &0.1300 & -0.01186(18) \\
			2&32 &64 &2.25 &0.1300 &  0.01391(17)\\\hline\hline
			3/2&24 &48 &1.7 &0.1340 & -0.00097(22) \\
			3/2&32 &64 &1.7 &0.1340 & -0.00052(11) \\\hline\hline
			1&24 &48 &1.75 &0.1663 & 0.02687(27) \\\hline\hline
			1/2&32 &64 &1.75 &0.1495 & 0.0055(12) \\\hline
	\end{tabular}}
	\caption{Summary of the simulation parameters considered in the analysis of adjoint QCD with different number of fermions $N_f$. The gauge coupling $g$ is related to $\beta=\frac{4}{g^2}$ and the bare fermion mass $m$ to $\kappa=\frac{1}{2(m+4)}$. Wilson fermions imply additive and multiplicative renormalization of the fermion mass, which means a more relevant estimate for the mass parameter is obtained from the partially conserved axial current (PCAC) relation. Three levels of stout smearing have been done for most of the simulation, except for SYM, where only one level has been considered. \label{tablenf2}}
\end{table}

%

\subsubsection{Parameter range and finite volume corrections}
The running of $\gamma_{PS}(\mu)$ obtained with the local current for all $\beta$ values and lattice sizes $L=24, 32$ at $N_f=2$ can be seen in Fig.~\ref{nf2finiteL}. For the final analysis mass and finite volume volume corrections have to be taken into account. We assume that the mass corrections are under control since we have chosen runs with $am_{PCAC}<0.04$. In our investigations of the mass spectrum, we have seen that finite size effects are the most important deformation at this parameter range. 
In order to reduce the finite size effects, we utilize the volume scaling formula \eqref{volumescaling}. We obtain the required input at $t'=s^{2}t$ by interpolating the jackknife samples with an exponential function.

\begin{figure}[htbp!]
	\centering
	\begin{minipage}[b]{\textwidth}
		\centering
\scalebox{0.9}{		\includegraphics[width=\textwidth]{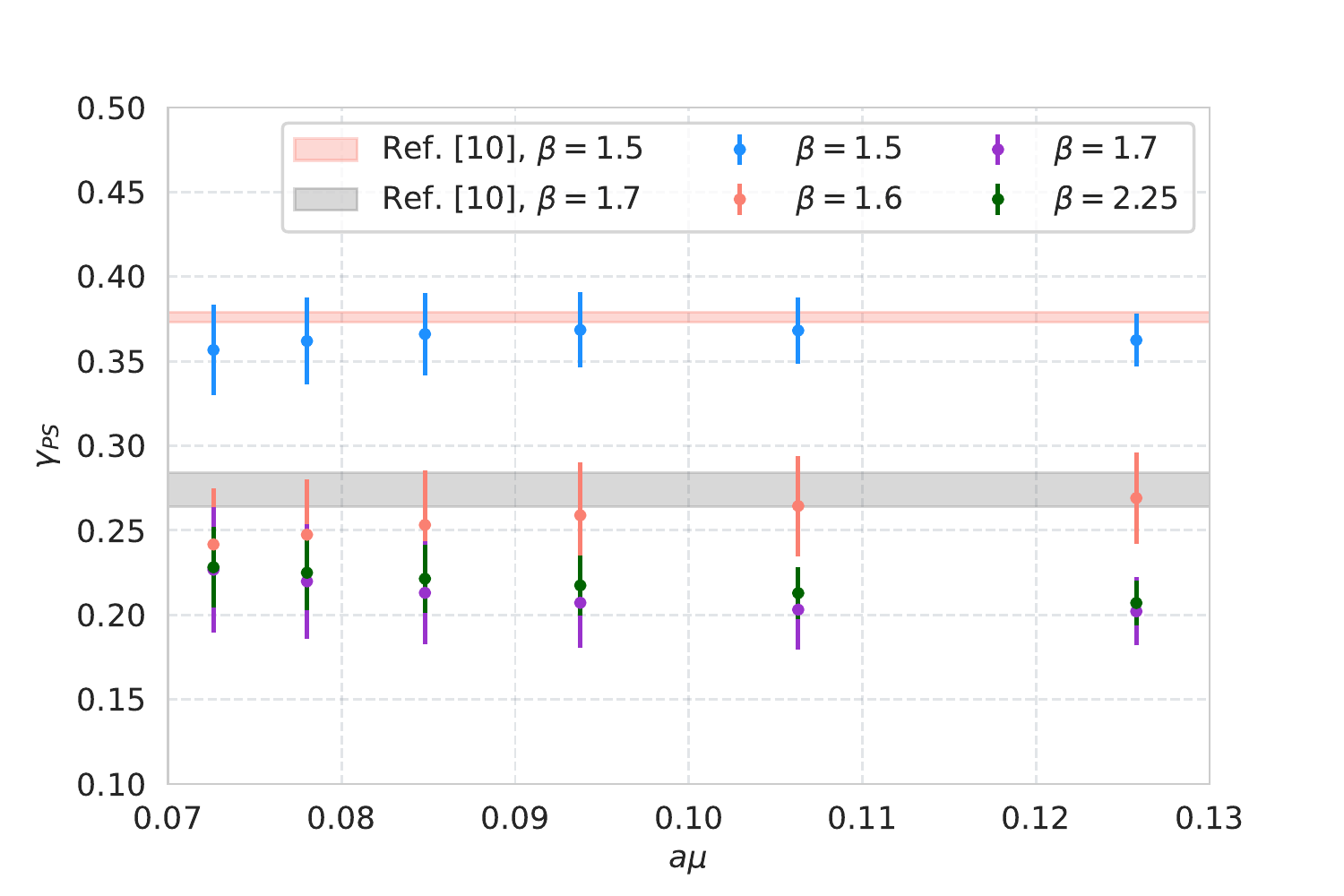}}
		\caption{$N_{f}=2$: $\gamma_{PS}(\mu)$ at $L=42$ for all available $\beta$ values (local vector current). For comparison, the red and gray bands show the results obtained from the mode number in Ref.\cite{Bergner:2016hip}.}
		\label{gammanf2}
	\end{minipage}
	\hfill

	\begin{minipage}[b]{\textwidth}
		\centering
	\scalebox{0.9}{	\includegraphics[width=\textwidth]{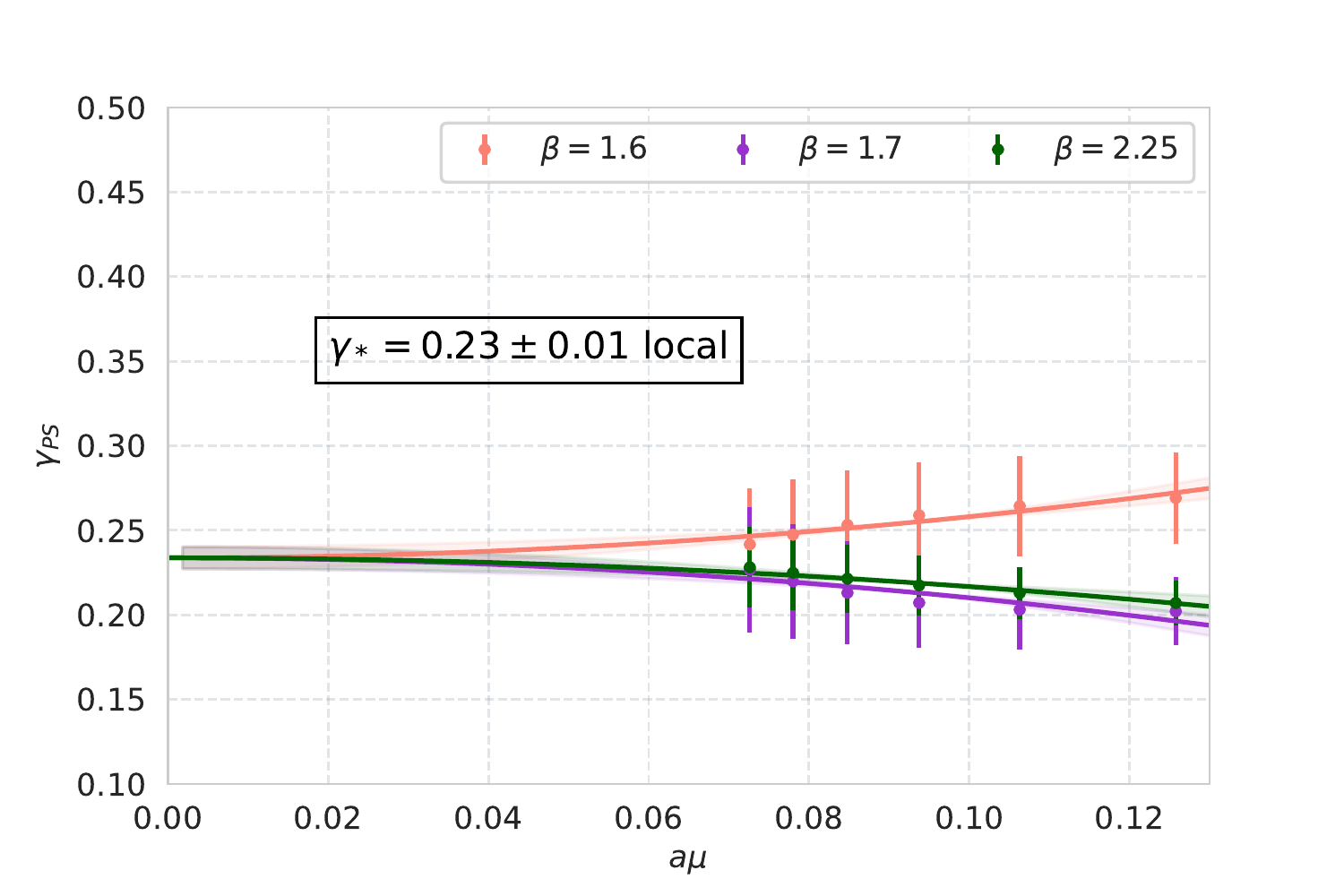}}
		\caption{$N_{f}=2$: Extrapolation of $\gamma_{m}$ to its critical value $\gamma_{*}$ at $\mu\to 0$ for the local vector current.}
		\label{gammanf2locextr}
	\end{minipage}
\end{figure}

After obtaining $\mathcal{R}_{PS}(s^{4}t)$ for $L=42$, we compute $\gamma_{PS}(\mu)$. In Fig.~\ref{gammanf2} we see that the $\beta=1.5$ results are not compatible with the existence of an universal $\gamma_*$ value in the deep infrared limit, as they don't overlap with the curves from the other three bare couplings. This is probably a result of additional systematic uncertainties due to the vicinity of the bulk transition and the larger finite mass corrections. Therefore, we concentrate on the $\beta=1.6$, $1.7$, and $2.25$ data. Fig.~\ref{gammanf2} shows that $\gamma_{PS}$, although not being strictly constant, shows a very weak scale dependence. This is a hint for an at least near conformal behavior.
 Moreover, the anomalous dimension of our larger $\beta$ values overlap at $a\mu\sim 0.08$. This behavior is in fact what we expect to see in a near conformal system. Indeed, in the limit $\mu\to 0$, all bare couplings must yield the same universal value $\gamma_{*}$.

\begin{figure}[htbp!]
			\begin{minipage}[b]{\textwidth}
						\centering
	\scalebox{0.9}{		\includegraphics[width=\textwidth]{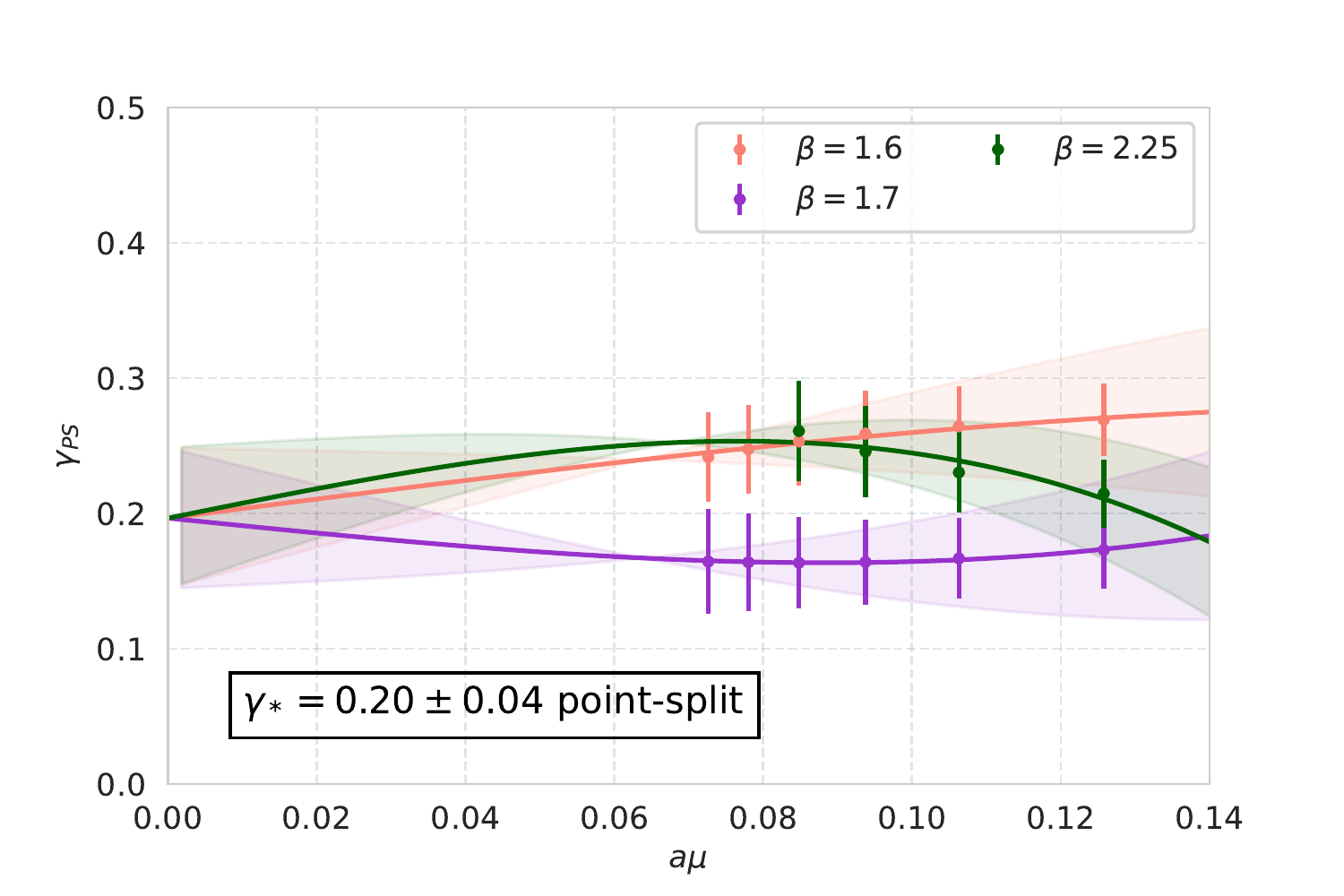}}
			\caption{$N_{f}=2$: Extrapolation of $\gamma_{PS}$ to its critical value $\gamma_{*}$ at $\mu\to 0$ for the point-split vector current.}
			\label{gammanf2nlextr}
		\end{minipage}
		\hfill
		
	 \begin{minipage}[b]{\textwidth}
	 			\centering
	\scalebox{0.9}{\includegraphics[width=\textwidth]{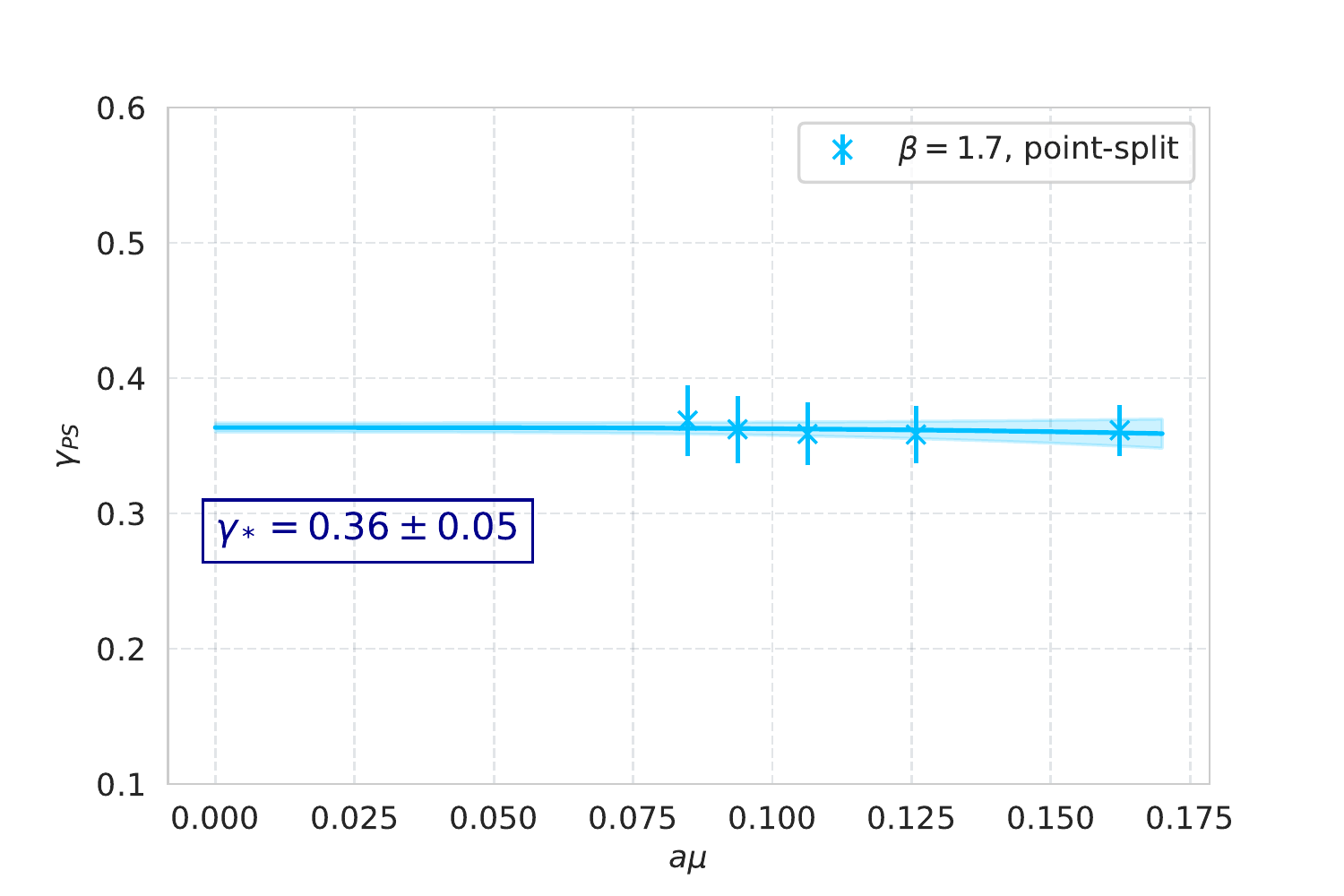}}
	\caption{$N_{f}=3/2$: Extrapolation of $\gamma_{PS}$ to its critical value $\gamma_{*}$ for the point-split vector current.}
	\label{gammanf3}
	\end{minipage}
\end{figure}

We present the final extrapolations towards the infrared limit in Figs.~\ref{gammanf2locextr} 
using the local current, while the point-split current results follow in Fig.~\ref{gammanf2nlextr}. 
To obtain the critical $\gamma_{*}$ we performed a global (joint) polynomial fit, where the parameter at $\mu=0$, i.e. $\gamma_{*}$, was common to all data sets. The remaining fit parameters are allowed to vary with $\beta$ and the uncertainties are obtained from the extrapolation of the jackknife samples. We have cut off the $\beta=2.25$ point-split data at $a\mu\sim 0.9$, since volume corrections start to become very large as $t$ increases. This fact is already observed in Fig.\ref{nf2finiteL}, where the values of $\gamma_{PS}(\mu)$ at $L=32$ and $L=24$ start to diverge from each other as the infrared is approached.  
We observe that the IR extrapolations from both local and point-split currents yield a compatible value of around $\gamma_*\approx 0.2$. It is smaller than previous estimations based on the mode number \cite{Patella:2012da,Bergner:2016hip}, but it is consistent with the Schrödinger-Functional results in \cite{Rantaharju:2015yva}. 

At this point, a short comment on $\beta=1.5$ is in order. Fig.~\ref{gammanf2} shows that at this $\beta$ value $\gamma_{PS}$ reaches a plateau at $\gamma_{*}\approx 0.38$, which is considerably larger than our extrapolated value. In our previous investigations, we have seen that a significantly larger mass anomalous dimension is obtained at this coupling. This value of the gauge coupling is just above the bulk transition for our tree level Symanzik improved gauge action and can in this sense be compared to the value of $\beta=2.25$ for the plain Wilson action used in \cite{Patella:2012da}. Indeed, a similar value for the mass anomalous dimension is obtained with the mode number in both cases \cite{Bergner:2016hip}. 
However, the large difference to the other $\beta$ value has to be taken into account. This will discussed in more detail in the following section.
Note that we have also considered possible phase transitions between $\beta=1.5$ and $\beta=1.6$, but we have found no evidence in the data of the temporal and spacial Polyakov line.

There is a rather large number of studies focused on the computation of $\gamma_{*}$ for $N_{f}=2$ adjoint QCD in the literature. Table \ref{tablenf2literature} summarizes our results and compare them to other studies found in the literature. From the Table it can be seen that our findings are compatible with many studies performed in the last years. Moreover, since we get a unique $\gamma_{*}$ value through the joint extrapolation, the gradient flow method might help to resolve the $\beta$ dependence of $\gamma_{*}$ in the deep infrared limit. Note that our previous data in \cite{Bergner:2016hip}, at $\beta=1.5$ and $1.7$ with the same lattice action have been $\gamma_{*}=0.376(3)$ and $\gamma_{*}=0.274(10)$ respectively. In comparison, our present study used much lighter fermion masses and an additional intermediate $\beta=1.6$, as well as a larger $\beta=2.25$.

\subsubsection{Running of the gradient flow coupling}

We can gain more information on the existence of a strong interacting IRFP by measuring the running of the renormalized gauge coupling. This can be computed through the gradient flow method, since the flowed gauge energy density is directly proportional to the renormalized coupling \cite{Luscher:2010iy}. There are two possible alternatives. The first possibility is to consider the running of the gauge coupling simply as a function of the scale $\mu$ proportional to $1/\sqrt{t}$

\begin{align}
	\label{gfcoupling}
	g^{2}_{\mathrm{GF}}(\mu)=\left. \frac{128\pi^{2}}{3(N^{2}-1)}t^{2}\langle E(t)\rangle\right|_{t=1/8\mu^{2}}.
\end{align}

\begin{figure}[htbp!]
	\centering
\scalebox{0.6}{\includegraphics{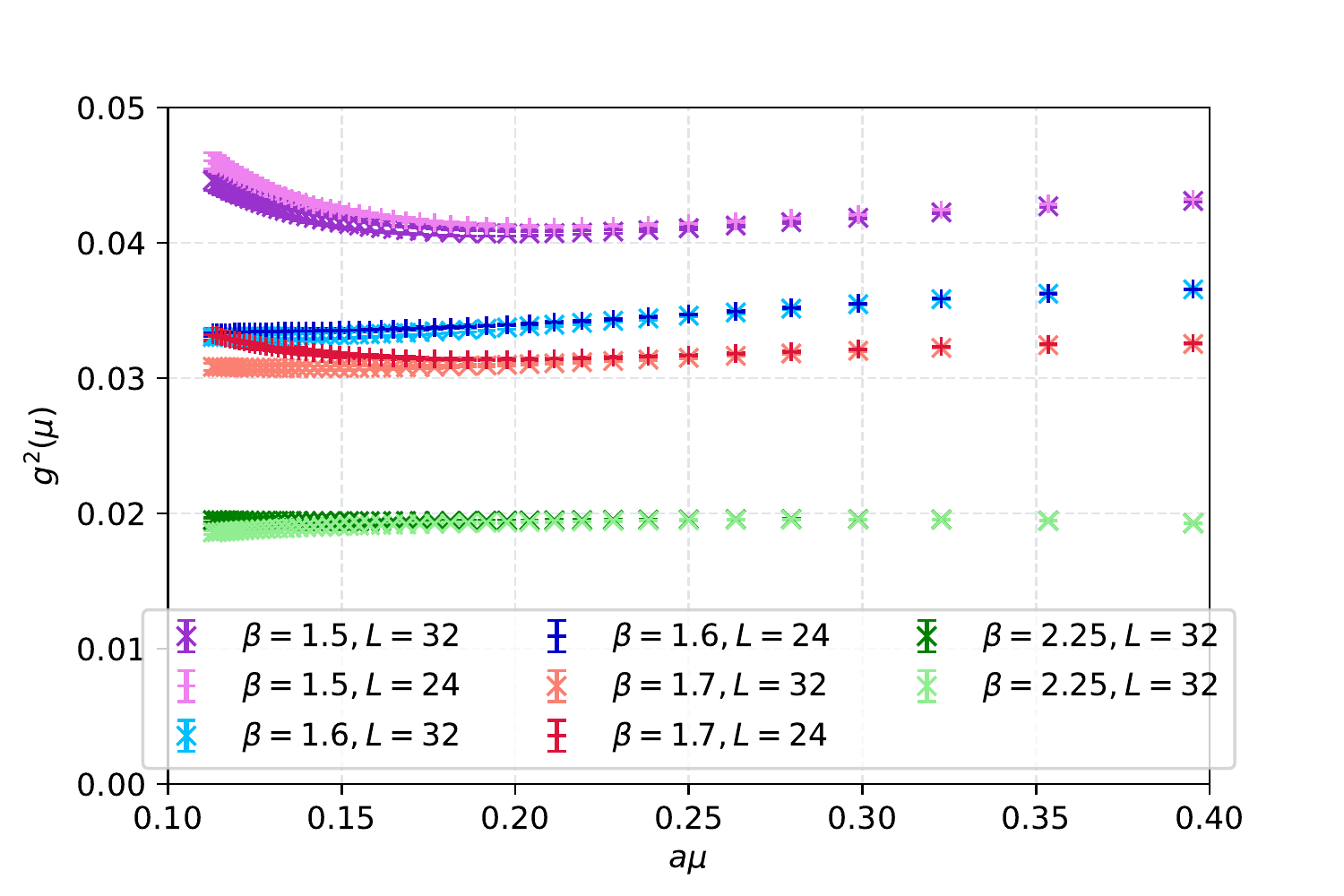}}
\caption{Gradient flow gauge coupling of $N_f=2$ adjoint QCD at different $\beta$ and volumes.\label{fig:coupling}}
\end{figure}

In this case, finite size and lattice spacing corrections must be addressed separately by extrapolating first the renormalized coupling to the infinite volume limit and then by sending the bare lattice gauge coupling to zero. This renormalized coupling is shown in Fig. \ref{fig:coupling}. The coupling shows a plateau at some intermediate scale, before it starts to run again at smaller $\mu$.

The second alternative allows for a better control on finite volume effects. Here one considers the running of the gauge coupling with respect to $\mu=1/L$ on the hypercubic torus of volume $L^4$ at fixed $c=\sqrt{8t}/L~$ \cite{Fodor:2012td}. The (discrete) $\beta$-function can be read directly from the difference of the renormalized coupling measured after flowing the gauge fields for a time equal to $t=\sqrt{c/L}$ for two lattices of size $L$ and $sL$

\begin{align}\beta(g^2)=\frac{g^2(sL)-g^2(L)}{\log{(s^2)}}\; ,\end{align}
where the step $s$ is customary chosen to be 2 or $3/2$. Here the constant $c$ defines the scheme, and has to be chosen in such a way that lattice discretization errors are under control. The continuum limit is reached in the limit $L/a \to\infty$ extrapolated for fixed renormalized coupling. When periodic boundary conditions are applied to gauge and fermion fields, as in our case, a finite volume correction has to be included in order to compute a correctly normalized coupling \cite{Fodor:2012td}. Given the limited range of bare gauge couplings of our simulations, we can compute the discrete $\beta$-function for only four points. Table~\ref{betaf} shows the results for the schemes $c=0.14, 0.20$ and $0.26$ with $s=32/24$. It can be seen that the $\beta$-function has a zero at $\beta=2.25$, while it becomes negative for $\beta=1.6$ and $1.7$. For larger $c$ the $\beta$-function at $\beta=1.5$ becomes positive, which can be understood as a consequence of the larger fermion mass bringing the system away from the fixed point.

\begin{table}[h]
	\centering
	\scalebox{0.9}{\begin{tabular}{ | c | c | c | c | }
			\hline
			$c$&$\beta$&$g^2$&$\beta(g^2)$\\\hline
			0.14 &2.25 &1.9E-2 &-8.0E-5  \\
			&1.7 &3.2E-2 &-1.6E-3  \\
			&1.6 &3.5E-2 &-2.4E-3  \\
			&1.5 &4.2E-2 &-2.3E-3  \\\hline
			0.20 &2.25 &1.9E-2 &1.3E-4  \\
			&1.7 &3.1E-2 &-1.1E-3  \\
			&1.6 &3.4E-2 &-1.4E-3  \\
			&1.5 &4.1E-2 &3.2E-4  \\\hline
			0.26 &2.25 &1.9E-2 &6.2E-4  \\
			&1.7 &3.1E-2 &-1.3E-3  \\
			&1.6 &3.3E-2 &-9.1E-4  \\
			&1.5 &4.2E-2 &3.2E-3  \\\hline
	\end{tabular}}
	\caption{Discrete $\beta$-function of $N_f=2$.\label{betaf}}
\end{table}

These results are compatible with the presence of an IRFP. In the near future we will complement this investigation of the $\beta$-function by considering more gauge couplings, what could serve to confirm that the theory is scale invariant in the infrared.
%
 By combining step-scaling together with the running of the mass anomalous dimension we could be able to extrapolate the value of $\gamma_*$ more reliably.

\begin{table}[h]
	\centering
	\begin{tabular}{ | c | c |}
		\hline
		&$\gamma_{*}$\\ \hline
		This study &\textit{local:}~~$0.23(1)$\\
		&\textit{point-split:}~~$0.20(4)$\\\hline
		Ref.  \cite{Bergner:2016hip} &$\beta=1.5:~~0.376(3)$\\
		&$\beta=1.7:~~~0.274(10)$ \\\hline
		Ref. \cite{Patella:2012da}& $0.371(20)$ \\ \hline
		Ref. \cite{Perez:2015yna}&$0.269(2)(5)$ \\\hline
		Ref. \cite{Rantaharju:2015yva} & $0.20(3)$  \\\hline
		Ref. \cite{DeGrand:2011qd}& $0.31(6)$ \\\hline
		Ref. \cite{DelDebbio:2010hx}& $0.22(6)$ \\\hline
		Ref. \cite{Giedt:2015alr}& $0.50(26)$\\\hline
		
	\end{tabular}
	\caption{Comparison of our results to values found in the literature for $N_{f}=2$.}  
	\label{tablenf2literature}
\end{table}

\subsection{Some estimations for smaller number of flavors}

We performed the same measurements for all the others lattices summarized in Table \ref{tablenf2}. The analysis was done in the same way as in the $N_{f}=2$ case. The only difference is that here we have notably less ensembles. In particular, for $N_{f}=3/2$ we only have one $\beta$ value. Although we employ the volume formula for this single $\beta$, we are of course not able to determine the $\beta$ dependence of $\gamma_{*}$, as we did in the previous section. The results for $N_{f}=3/2$ can be seen in Fig. \ref{gammanf3}. We clearly see that the anomalous dimension is almost a constant. This is a signal for the IR (near-)conformal behaviour of the theory. Hence, the system appears to lie inside the conformal window. We extrapolate to $\mu\to 0$ and obtain $\gamma_{*}=0.38(2)$. As expected, the value of $\gamma_{*}$ is larger than in the two-flavour case. Our result is in agreement with previous lattice investigations found in Refs.~\cite{Bergner:2017gzw,Bergner:2017bky}. Especially, we get the same value as in Ref. \cite{Bergner:2017gzw}, where the authors found $\gamma_{*}\sim 0.38$ based on an analysis of the mode number.

\begin{figure}[htbp!]
	\centering
	\scalebox{0.9}{ \includegraphics[width=\textwidth]{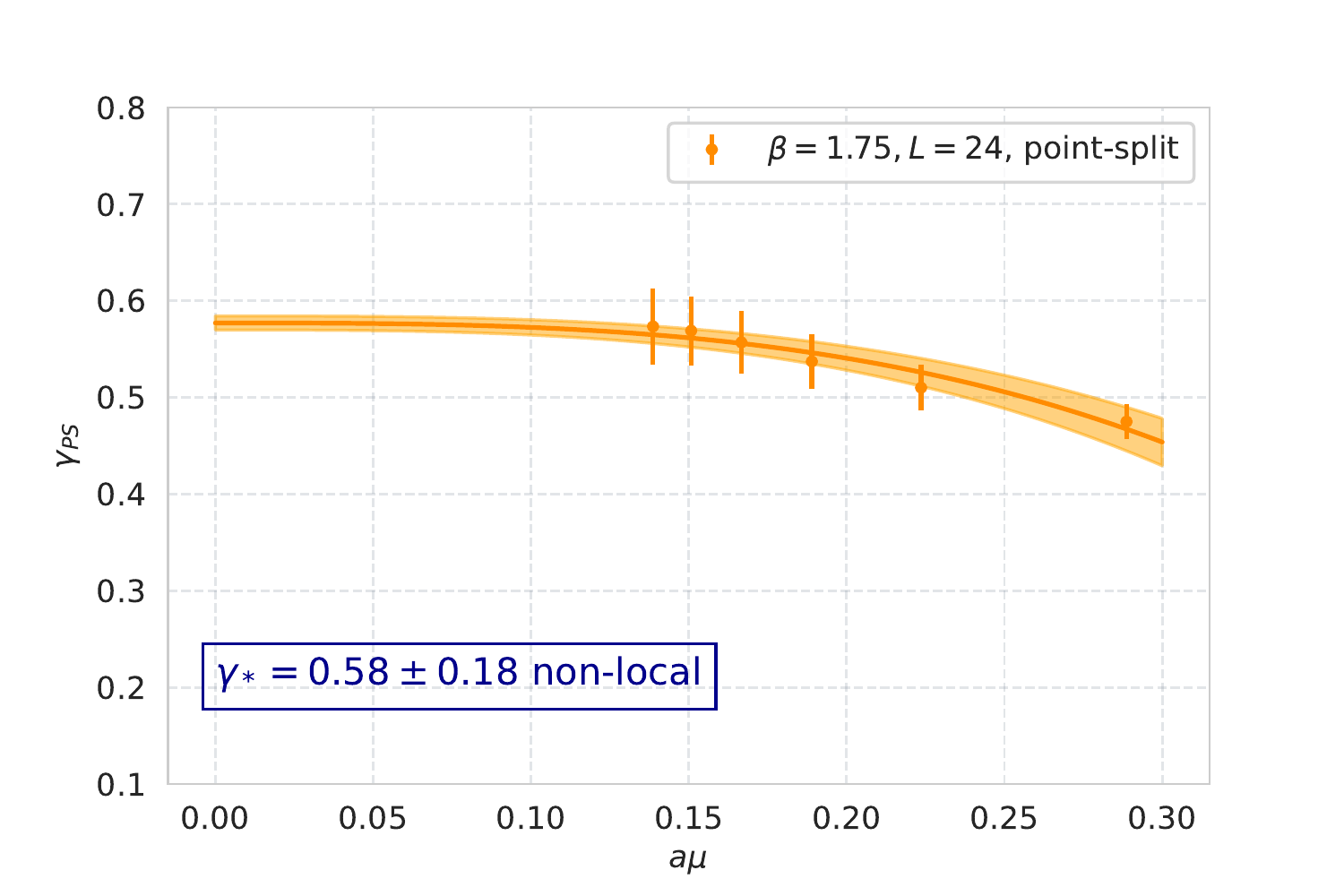}}
	\caption{Extrapolation of $\gamma_{PS}$ to its critical value $\gamma_{*}$ for $N_f=1$ using the point-split current.}
	\label{gammanf1}
\end{figure}

In the one-flavour case we only have one $\beta$ and one volume. The results are shown in Fig. \ref{gammanf1}. Although for these parameters we are already able to see a (near-)conformal behaviour, i.e. a very weak change in $\gamma_{PS}$, the extrapolated value $\gamma_{*}$ should be taken very carefully. In particular, it is considerably smaller than the results in Ref. \cite{Athenodorou:2014eua}, which estimated $\gamma_{*}\sim 0.9$. It is however a valuable piece of information to see that it is likely for the theory to lie in or very near the conformal window. The smaller values of $\gamma$ indicate rather a conformal scenario.

\begin{figure}[htbp!]
	\centering
	
	
	\begin{minipage}[b]{0.49\textwidth}
		\includegraphics[width=\textwidth]{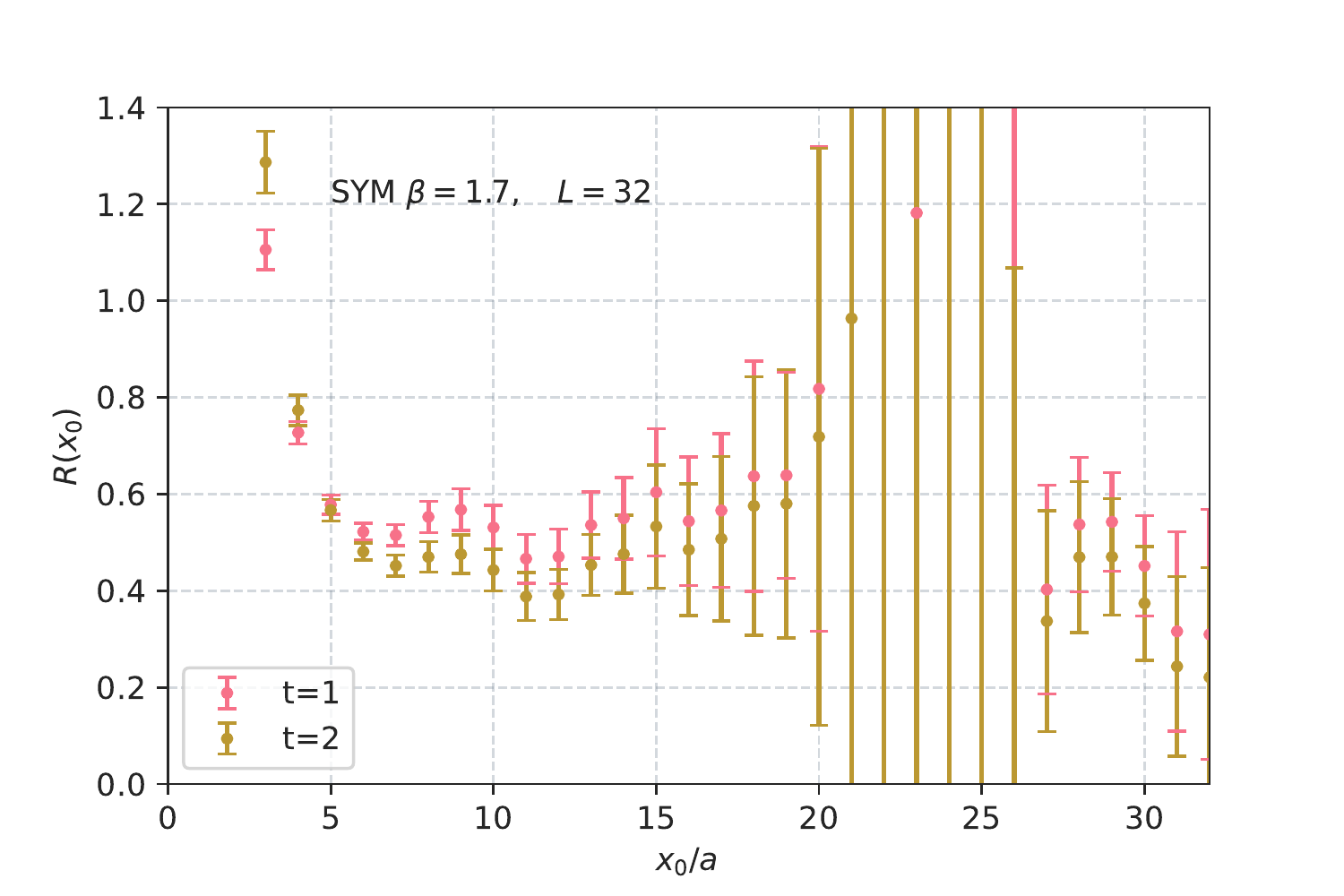}
	\end{minipage}
	\hfill
	\begin{minipage}[b]{0.49\textwidth}
		\includegraphics[width=\textwidth]{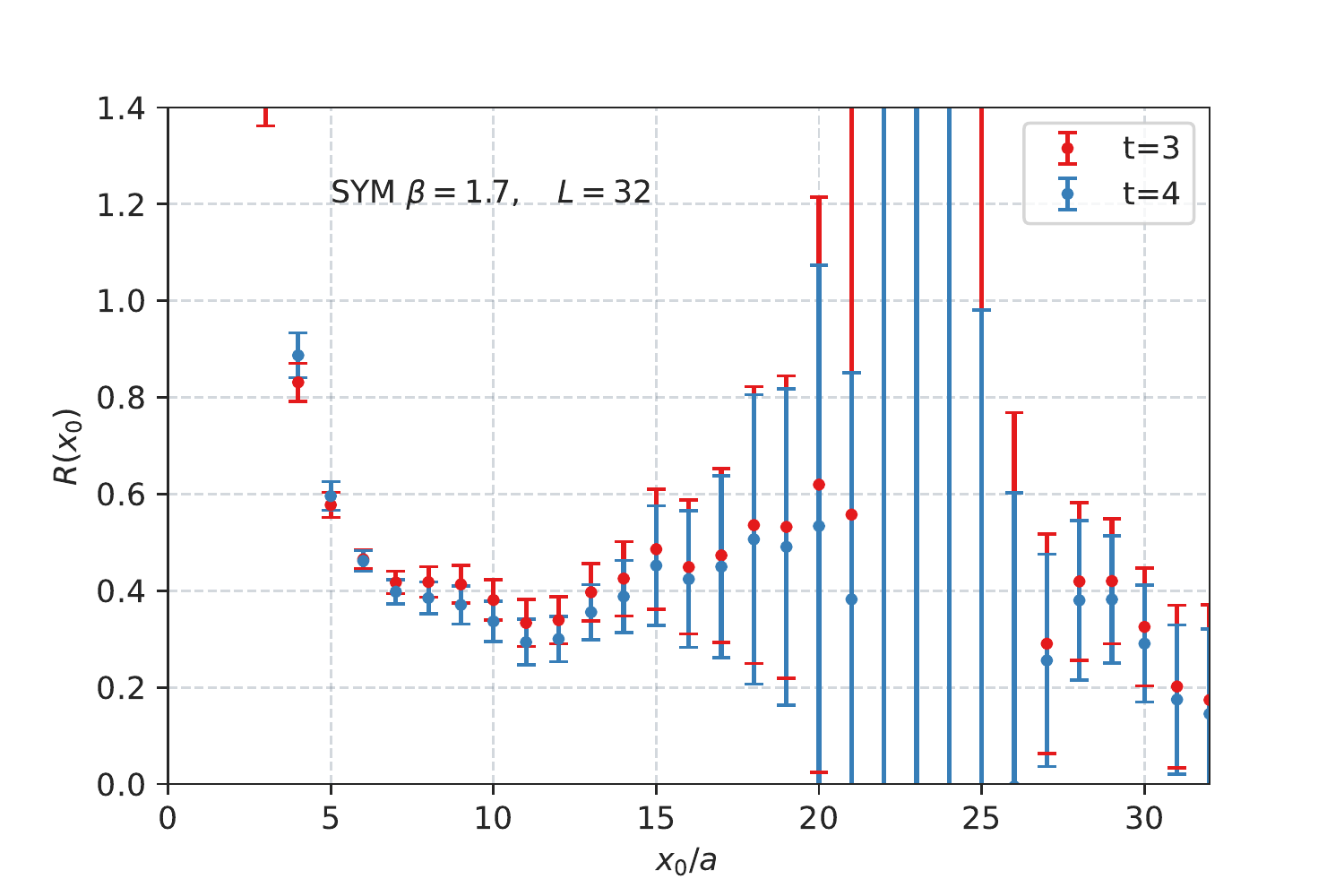}
	\end{minipage}
	\caption{$\mathcal{R}(x_{0})$ in SU(2) SYM (local vector current). Large errors lead to an unreliable signal at large $x_{0}$.}
	\label{Rsym}
	
\end{figure}

\begin{figure}[htbp!]
	\centering
	\scalebox{0.8}{\includegraphics[width=\textwidth]{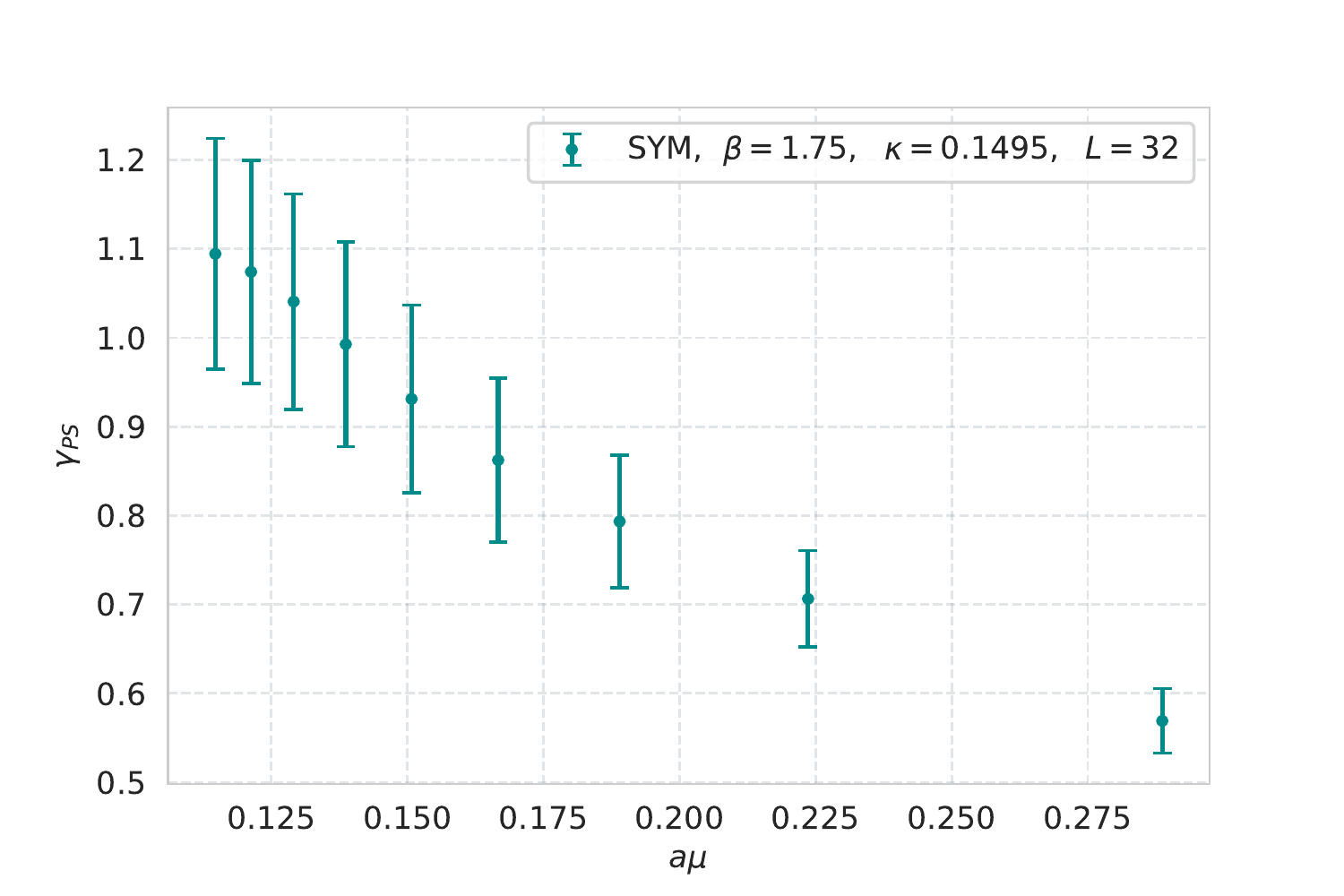}}
	\caption{Scale dependence of $\gamma_{PS}$ in SU(2) SYM using the local vector current.}
	\label{gammasym}
\end{figure}

Finally, in SYM, as expected, we don't see any freezing in the running of $\gamma_{m}$. This can be seen in Fig.~\ref{gammasym}. From Fig. \ref{Rsym} we see that the estimation of the plateau of $\mathcal{R}(x_{0})$ is particularly difficult. If one insists in averaging over some $x_{0}$ interval, the resulting $\mathcal{R}(t)$ yields a $\gamma_{PS}(\mu)$ that shows a strong scale dependence in the whole considered range of $\mu$. This result is indeed a signal for a system lying well below the lower edge of the conformal window, as predicted for SYM. In general, it is expected that the scale dependence becomes much stronger in a chiral broken theory, compared to the conformal case.


\section{Conclusions}

We have studied the mass anomalous dimension for different theories 
with fermions in the adjoint representation using a recently proposed method
based on the Gradient flow. We have started the analysis with the theory 
with $N_f=2$ Dirac flavors.
Previous lattice investigations of this theory established strong indications for 
a conformal behavior. 
However, the predictions for the mass anomalous dimension differ significantly. 
In particular, different values of the coupling constant lead to
different predictions. Our aim is to test the Gradient flow method and resolve 
these discrepancies.

The results resemble some features of our investigations with other methods, 
in which the smallest $\beta$ value favors a larger mass anomalous 
dimension. However, since the method is able to resolve the scale dependence 
of the running of the renormalized mass in a certain region of the 
renormalization group flow, we can extrapolate results from different gauge 
couplings to a prediction for the infrared fixed point. This leads 
to a smaller value of the mass anomalous dimension in the range of 
$\gamma_{*}\approx 0.16$ to $0.24$. 

We have investigated the dependence of the mass anomalous dimension at 
the fixed point obtained by the gradient flow method on the
number of fermion flavors. The first observation is a change of the $\gamma(\mu)$ 
dependence on the flow scale $\mu$. As expected, no fixed point 
value can be extrapolated in the SYM theory ($N_f=1/2$). 
The observed scale dependence gets weaker for larger $N_f$. 
We find that the extrapolated value increases for smaller $N_f$. 
The largest value is $\gamma=0.58(18)$ for $N_f=1$ which decreases to 
$\gamma=0.36(5)$ for $N_f=3/2$. These results give the first reasonable estimates 
from the gradient flow method, but further runs with different gauge
couplings would be required for a more detailed and well established
picture.
\section{Acknowledgements}
We thank Anna Hasenfratz, Ethan Neil, and Gernot Münster for helpful discussions.
The authors gratefully acknowledge the Gauss Centre for Supercomputing
e.~V.\,\linebreak(www.gauss-centre.eu) for funding this project by providing
computing time on the GCS Supercomputers JUQUEEN, JURECA, and JUWELS at J\"ulich Supercomputing
Centre (JSC) and SuperMUC at Leibniz Supercomputing Centre (LRZ). This work is
supported by the Deutsche Forschungsgemeinschaft (DFG) through the Research
Training Group ``GRK 2149: Strong and Weak Interactions - from Hadrons to
Dark Matter''. G.~Bergner acknowledges support from the Deutsche
Forschungsgemeinschaft (DFG) Grant No.\ BE 5942/2-1. This work was supported in part by the U.S. Department of Energy, Office of Science, Office of High Energy Physics under Award Number DE-SC-0010005 (C.L.)


\end{document}